\documentclass[%
 reprint,
 amsmath,amssymb,
 aps,
]{revtex4-2}
\usepackage[usenames,dvipsnames]{xcolor}
\usepackage{amssymb}
\usepackage{siunitx}
\usepackage{comment}
\usepackage{hyperref}
\newcommand{\um}{\,\si{\micro \meter}}
\newcommand{\kB}{k_\mathrm{B}}
\newcommand{\pard}[2]{\frac{\partial{#1}}{\partial{#2}}}

\usepackage{graphicx}
\usepackage{dcolumn}
\usepackage{bm}%
\usepackage{ulem}
\newcommand{\JP}[1]{\textcolor{blue}{#1}}

\begin{document}

\preprint{APS/123-QED}

\title{Huygens' clocks at the microscale}

\author{Yaocheng Li$^1$}
\author{Ivan Palaia$^2$}
\author{Jinzi Mac Huang$^3$}
\author{Antoine Aubret$^4$}
\author{Jérémie Palacci$^{5}$}


 \email{jeremie.palacci@ist.ac.at}
\affiliation{$^1$Department of Physics, University of Chicago, USA}
\affiliation{$^2$Department of Physics, King’s College London, London, United Kingdom}
\affiliation{$^3$New York University Shanghai, China}
\affiliation{$^4$LOMA, CNRS, UMR 5798, Talence, France}
\affiliation{$^5$Institute of Science and Technology Austria, Klosterneuburg, Austria}


\begin{abstract}
Weakly coupled oscillators adjust their dynamics to work in unison: they synchronize. This ubiquitous phenomenon is observed for oscillating pendulum, electronic devices, as well as clapping crowds or flashing fireflies. In effect, synchronization constitutes an efficient mean to translate microscopic into large scale dynamics. Synchronization has been broadly studied theoretically with  most notable experimental results at the microscale on the pattern of beating flagella coupled hydrodynamically. Studies on minimal platforms leveraging synthetic colloidal platforms remain however limited. 
Here we devise and study a model system of noisy and "measurably imperfect" colloidal oscillators: autonomous clocks made of an active swimmer revolving around a passive sphere and coupled chemically. The distribution of natural frequency of the clock is achieved using passive spheres of various sizes, thus without altering the (phoretic) coupling  between clocks. We observe that pairs of oscillators lock phases before slipping and returning to sync, and we characterize the synchronicity of the pair. We rationalize our findings with a stochastic model, formalizing synchronization  as a classical Kramers escape problem, in quantitative agreement with the experiment. Our results set a blueprint for synchronization with micrometric autonomous systems.
\end{abstract}

\maketitle

{\bf Introduction.}
Christiaan Huygens originally reported that clocks hung together on a heavy beam  eventually oscillate periodically in unison: they synchronize~\cite{sprat1722history,Huygens1888_OeuvresCompletes} 
The weak interaction mediated by the beam was key to the phenomenon, highlighting  the role of adequate coupling in synchronization~\cite{bennett2002huygens}. Synchronization is  ubiquitous, observed across scales and objects with interactions of varying complexity.
Interacting metronomes~\cite{pantaleone2002synchronization}, arrays of pendulums~\cite{wiesenfeld2014spontaneous}, walking pedestrians on a bridge~\cite{fujino1993synchronization,fujino2016conceptual}, or superconducting Josephson junctions~\cite{wiesenfeld1996synchronization} synchronize.
Synchronization is also widespread in complex and biological systems: population dynamics~\cite{marlatt1907periodical,post2002synchronization}, locomotion of microorganisms~\cite{vilfan2006hydrodynamic,wan2016coordinated,polin2009chlamydomonas}, clapping of audiences  \cite{neda2000selfsorganizing}, and bodily functions~\cite{breunig2010cilia,kozlov2011forces}, notably underpinning the emergence of metachronal waves~\cite{brumley2015metachronal}.
From a practical perspective, synchronization constitutes a robust method to propagate the dynamics of animate constituents across length scales and give rise to emergent macroscopic dynamics.

The investigation of synchronization at the microscale, where population variability and 
noise exist, remains challenging. Extensive studies with {\it Chlamydomonas} algae as model system have shown complex dynamics of synchronization arising from noise, including synchronization and slips from beating flagella  with hydrodynamic coupling~\cite{polin2009chlamydomonas, eLife.Brumley.2014, Goldstein:2009du, Goldstein2011, Leptos2013}.  Experiments with minimal,  synthetic systems have however been limited to optically driven, phase-free colloids also coupled hydrodynamically~\cite{damet2012hydrodynamically,juniper2017dynamic,bruot2011noise,bruot2012driving,kotar2010hydrodynamic}. 
In this work, we leverage tools of active matter to devise colloidal clocks and study the synchronization of autonomous oscillators at the microscale and coupled chemically rather than hydrodynamically. The clocks are assembled from an active self-propelled particle (the ``hand'') revolving around a passive sphere.
They are autonomous and phase-free, the hand rotating in the absence of external drive. They exhibit fluctuations in the hand motion and natural frequency polydispersity.

\begin{figure}[htbp]
\centering
\includegraphics[width=0.9\linewidth]{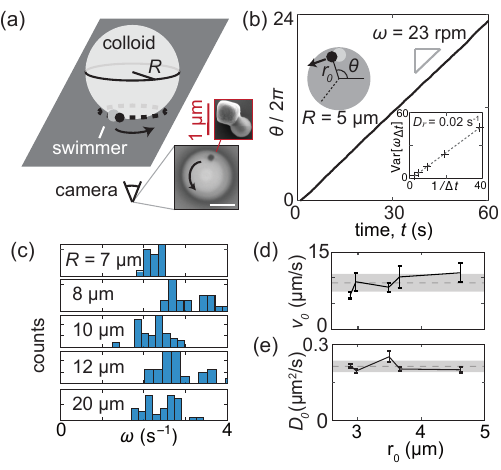}
\caption{{\bf Colloidal clocks}.
(a) Scheme of the colloidal clock. A colloidal microswimmer (inset) revolves around a passive sphere of radius $R$. Scale bar is 5\um.
(b) Normalized cumulated phase angle $\theta / 2 \pi$ for a single colloidal clock with passive colloid radius $R=\SI{5}{\micro m}$, as a function of time.
The slope provides the natural frequency of the clock, $\omega$, with high accuracy.
(b-inset) The variance $\mathrm{Var}[\omega_{\Delta t}]\propto D_r/\Delta t$ quantifies the rotational noise $D_r=0.02~\mathrm{s^{-1}}$ for the  clock. 
(c) Distribution of natural frequencies for this study obtained by changing the size $R=7, 8, 10, 12, 20$\um~of the central sphere.
(d) Constant tangential speed $v_0 = \omega r_0$ for colloidal clocks of  radii $r_0 \approx \sqrt{2 R d_s}$. (e) Constant translational diffusion $D_0$ for colloidal clocks of  radii $r_0 \approx \sqrt{2 R d_s}$.}
\label{fig1}
\end{figure}

{\bf Assembly of a colloidal clock.} The hand of the clock is an active colloid, $d_{\mathrm{s}}\sim  1.5\um$ in size, made of a passive polymer (TPM)  and a photocalytic (black) hematite heading the propulsion. This particle is a known colloidal microswimmer, fueled in hydrogen peroxide solution, whose direction of propulsion and sign of diffusiophoretic interactions can be controlled by changing the pH of the solution~\cite{Palacci:2014bt,Aubret:2018cca,Anderson}.
At the considered basic conditions, $\mathrm{pH}=9$, the hematite heads the microswimmer and diffusiophoretic interactions are  attractive.
Large passive beads (with radii $R=5-20\um$) are added to the capillary and confined by gravity to the bottom substrate of the capillary, where observations are carried [SI Sec.~S1].
A passive bead is trapped using custom-built optical tweezers and manually displaced towards a self-propelled colloid. Upon close-approach and collision, the self-propelled colloid is captured by the sphere, as previously reported for active particles near boundaries~\cite{Takagi:2014hf,Simmchen:1iy, 10.1038/s41467-022-29430-1}.
The active colloid revolves around the passive sphere, wedged between its perimeter and the bottom substrate [Fig.~\ref{fig1}(a)]. The combination of the passive sphere and the revolving microswimmer constitutes the colloidal clock, with the active colloid being the hand. The revolution is  persistent, unidirectional and subject to noise, the result of fluctuations of propulsion orientation.  The directions of revolution, clockwise or counterclockwise, are equivalent and set by the initial capture of the active colloid by the passive sphere. In rare instances, the noise results in the release of the self-propelled particle from the surface of the sphere or the reversal of the directions of revolutions, leading to the breakdown of the clock.
We only consider clocks that are stable for more than 10 minutes and hundreds of revolutions. The intrinsic frequency of the clock is set by the revolution rate of the active particle around the sphere, away from other clocks.
For example, the cumulated angle $\theta$ defined by the hand linearly increases with time during the $10$ minutes of an experiment [Fig.~\ref{fig1}(b)], allowing us to define the system as a clock with constant pulsation $\omega$ with high accuracy (here, $\omega = 23.0\,\mathrm{rpm} = \SI{2.41}{s^{-1}}$). 
Intuitively, the revolution of the colloid arises from the self-propulsion around a circumference of radius $r_0 \approx \sqrt{2 R d_s}$ [sketch in Fig.~\ref{fig1}(b) and SI Sec.~S4]: this sets $\omega = v_0/r_0$. The tangential velocity of the hand $v_0 = (9.0 \pm  1.5)\, \si{\micro m /s}$ [Fig.~\ref{fig1}(d)] is lower than the propulsion velocity $\sim 12$\,\um/s of a  microswimmer in free space, and independent of the passive sphere radius. 
Repeating this assembly, we devise a variety of microscopic and autonomous clocks with stable frequencies $\omega = v_0/\sqrt{2Rd_s}$,  by changing the radius $R$ of the passive sphere [Fig.~\ref{fig1}(c)].
This enables us to control the frequency of the clocks while keeping fixed the non-equilibrium parameters and the coupling between clocks and, importantly, making it possible to study synchronization in well-controlled settings. It  contrasts with approaches where frequency is set by varying the speed of the microswimmers through changes of hydrogen peroxide concentration or light intensity, which inevitably lead to changes in the coupling between the active particles.
We next define the instantaneous angular velocity ${\omega}_{\Delta t}(t) = [\theta(t+\Delta t) - \theta(t)]/\Delta t$ obtained by tracking the (black) hematite part of the active particles over a time $\Delta t$. Its  variance 
$\mathrm{Var}[\omega_{\Delta t}]\propto D_r/\Delta t$ [Fig.~\ref{fig1}(b) inset] provides a quantitative estimate of the rotational noise of the clock, $D_r = \SI{0.02} {s^{-1}}$. We repeat the measurement for clocks of different radii $r_0$ and observe that the translational diffusion coefficient $D_0= r_0^2 D_r=  (0.22 \pm 0.02)$\,\um$^2$/s, is constant  [Fig.~\ref{fig1}(e)] and comparable to the thermal  diffusivity of particles of same size, suggesting thermal (rather than active) fluctuations are the source of noise. 

The active colloid is powered by decomposition of hydrogen peroxide, which diffuses rapidly in comparison with the time it takes for the active colloid to travel. The result is a so-called low-P\'eclet dynamics, $Pe=r_0^2 \omega/D_\mathrm{chem}\sim 0.01 \ll 1$, where $D_\mathrm{chem}\sim 10^3$\um${}^2/\mathrm{s}$ is the diffusivity of the chemicals.
In the (moving) reference frame of the revolving active colloid, this leads to a \textit{steady} chemical concentration $c\propto 1/r$, solution of $\Delta c=0$.
Hence, neighboring particles experience phoretic attraction with an effective potential $V=-k/r$, where $k>0$ characterizes the diffusiophoretic attraction,  $k\sim 80\,\kB T {\cdot} \um$ in comparable conditions \cite{10.1021/acsnano.4c18078}. 
Taken together, we devised a model system of  autonomous colloidal clocks, whose frequency can be controlled with high precision and independently from the non-equilibrium coupling---a long-range attractive potential $\propto 1/r$ mediated by chemical gradients. 

\begin{figure}
\centering
\includegraphics[width=0.9\linewidth]{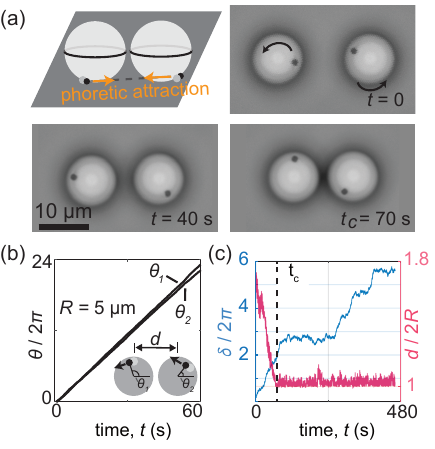}
\caption{{\bf Pairs of colloidal clocks.}
(a) Timelapse formation of a pair of  co-rotating colloidal clocks. Clocks are assembled initially, far apart (not presented). They are approached using optical traps and released ($t = 0\,$s). They then attract phoretically until  contact ($t_\mathrm{c}=70\,$s). The equators remain in contact and the distance of the centers of the colloidal clocks is constant $d=2R$.
(b) Cumulated angle of rotating clocks  before they come in contact. The slope is the natural frequency, that appear close but not identical.
(c) Dynamics of the relative phase $\delta$. Delta increases linearly before contact $t_\mathrm{c}$: the clocks are not synchronized. After $t_\mathrm{c}$, $\delta$ shows plateaus of phase-locking  near odd values of $\pi$ indicative of synchronicity. 
}
\label{fig2}
\end{figure}

{\bf Synchronization  of colloidal clocks.} Multiple colloidal clocks are assembled and their natural frequencies and directions of rotation measured when distant from each other [Fig.~\ref{fig2}(a)]. Slight differences in the size of the clock or between swimmers lead to small differences in intrinsic clock frequency [Fig.~\ref{fig1}(c)], as visible from the differences of cumulative angles of each clock [Fig.~\ref{fig2}(b)]. We leverage this polydispersity to investigate the effect of frequency variability on the synchronization of clocks at the microscale. Distant clocks are approached with optical traps,   subsequently removed, leaving clocks free.
The two clocks then continue approaching as a result of diffusiophoretic attraction, until the passive beads make contact at the equator at time $t_\mathrm{c}$ [Movie S1 and S2, Fig.~\ref{fig2}(a, c)]. Pairs that revolve in the same direction constitute a system of co-rotors, pairs with opposite direction of revolution are counter-rotors.

\begin{figure}
\centering
\includegraphics[width=\linewidth]{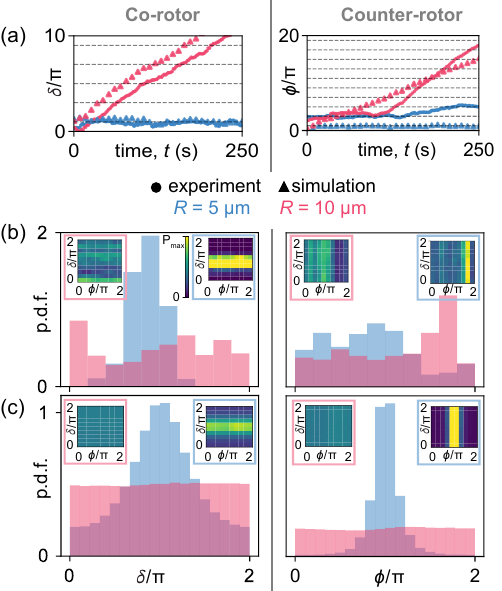}
\caption{\textbf{Synchronization of colloidal clocks.}
(a) Temporal dynamics of the phases for co-rotor and counter-rotor clock pairs. For both experimental and simulation results, plateaus are manifested for the clock pair with smaller colloid radius (blue), suggesting strong synchronization. Conversely, the pair with larger radius (pink) manifest phase slip, pointing at little to no synchronization.
(b) Experimental and (c) simulated probability density function (p.d.f.) of the relative phases, $\delta$ for co-rotors and $\phi$ for counter-rotors, for the clock pairs from panel (a). Insets contain the joint distributions for large (left, pink) and smaller (right, blue) passive colloids.
} 
\label{fig3}
\end{figure}

In order to study their synchronization, we compare the relative phases of pairs of clocks, and compute the difference between the cumulative angles $\delta = \theta_2 - \theta_1$ for co-rotating clocks, and their sum $\phi = \theta_1+\theta_2$ for counter-rotating clocks. We observe that $\delta$ increases monotonously before $t_\mathrm{c}$ [Fig.~\ref{fig2}(b)-(c)], indicating that before contact, clocks revolve independently at their natural frequency. 
From the instant of contact, $t=t_\mathrm{c}$, the clocks remain in contact with their centers at distance $d=2R$ [Fig.~\ref{fig2}(b)], while active colloids keep revolving.
Remarkably, the contact is swiftly followed by phase-locking between clocks, visible by the formation of plateaus in $\delta$ for corotating clocks and indicative of synchronization: synchronous states are interrupted by drifting states, terminated by the reformation of plateaus [Fig.~\ref{fig2}(c)].
Comparable dynamics is observed for counter-rotating pairs, with $\phi$ drifting before contact and synchronizing afterwards [SI Sec.~S4]. The temporal dynamics of the phases for synchronised co- or counter-rotating clocks are represented in Fig.~\ref{fig3}(a) in blue. 
The phase angle ($\delta$ for co-rotors, $\phi$ for counter-rotors) exhibits plateaus of synchronization nearing $(2n+1)\pi$, where $n$ is an integer. 
The probability distributions for $\theta$ and $\phi$, whose values we now reduce to the interval [0, $2\pi$), show a peak near $\pi$, more pronounced for $\delta$ in counter-rotors than for $\phi$ in co-rotors [Fig.~\ref{fig3}(b)].
The synchronization of colloidal clocks is similarly visible in the joint distribution of $(\phi,\delta)$, where high probability density appears as a bright strip near $\phi= \pi$ for counter-rotors and near $\delta= \pi$ for co-rotors [Fig.~\ref{fig3}(b) insets]. 

\begin{figure}[htbp]
\centering
\includegraphics[width=0.9\linewidth]{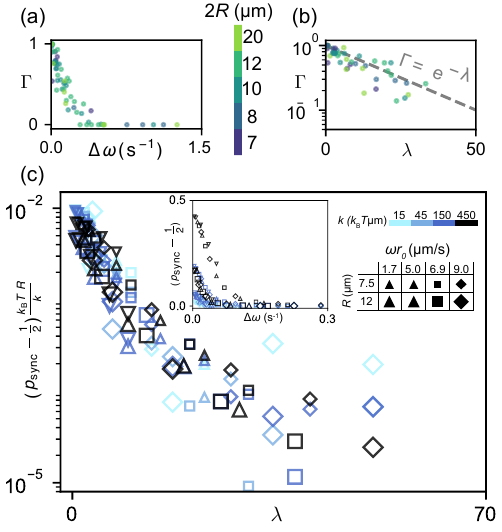}
\caption{\textbf{Synchronization  for co-rotors with different colloidal radii.}
(a) Synchronization ratio $\Gamma(\Delta \omega)$ between two colloidal clocks vanishing for  $\Delta \omega\gtrsim0.3~\mathrm{s^{-1}}$.
(b) Collapse of the experimental data onto the theoretical prediction $\Gamma=\exp(-\lambda)$ without fitting parameter (dashed line).
(c) Simulation data of probability of being in a synchronized state for a broad range of attractive coupling strengths $k$, swimming speed $\omega r_0$, and radii $R$, collapsed  as predicted by theoretical modeling. (inset) Simulation data prior to collapse.
}
\label{fig4}
\end{figure}

 Experiments with clocks of larger radii show limited synchronization, resulting from reduced coupling as their hands are on average more distant,as exemplified for clocks with 10 $\mu$m vs 5 $\mu$m passive sphere radii [Fig.~\ref{fig3}(a-b), pink].
Incidentally, experiments in seemingly identical configurations could lead to shorter plateaus and poor synchronization, which we intuited resulted from intrinsic noise ({\it i.e.}~small measured differences of natural frequencies of the clocks).
In order to quantify our observation, we define $\Gamma = t_{s}/\tau$, the ratio  between the time $t_s$ spent by two clocks in sync, {\it i.e.}~the sum of the duration of all the plateaus, and the total duration of experiment \JP{$\tau$}.
$\Gamma$ constitutes an order parameter for synchronization, monotonically decreasing from $1$ for a perfectly synchronized pair to $0$ for an asynchronous one.
Experimental measures of $\Gamma$ for our experiments with counter- or co-rotating clocks of different sizes and/or slightly different natural frequency, over 50 pairs are gathered in Fig.~\ref{fig4}(a). 
The data collapse when plotted against $\Delta \omega = \big\vert |\omega_1|-|\omega_2|\big\vert $, the difference of natural frequencies for the pair, without notable differences of synchronization between co-rotating and counter-rotating pairs [SI Sec.~S4], or effect of the size of the colloidal clock (hence distance between clocks and coupling strength).
In addition, the synchronicity rapidly decreases and vanishes beyond $\Delta \omega \sim 0.3\,\si{s^{-1}}$, highlighting the role of fluctuations  and intrinsic noise (polydispersity in natural frequencies) for slipping between synchronous states. 

{\bf Theoretical modeling.} To model the experiment, we study  a pair of autonomous oscillators with natural frequencies $\omega_{1,2}$, subject to noise and coupled by an attraction $\propto 1/d_{12}$, where $d_{12}$ is the distance between the two active colloids. The dynamical equations for phase $\delta$ and $\phi$ are obtained from linear combinations of the dynamical equations for $\theta_{1,2}$ [SI Sec.~S2], leading to: 
\begin{eqnarray}
    \dot{\delta } &=& (\omega_2-\omega_1) - \mu_0 \pard{V}{\delta}+ 2 \sqrt{D_r}\,\xi_\delta
    \label{eq:deltadotgen}
    \\
    \dot{\phi } &=&   (\omega_1+\omega_2) - \mu_0 \pard{V}{\phi} + 2 \sqrt{D_r}\,\xi_\phi \,,
    \label{eq:phidotgen}
\end{eqnarray}
where $\omega_1$ and $\omega_2$ are the natural frequency of each clock, $\mu_0$ is the mobility (assumed equal for the two microswimmers), $D_r = D_0 / r_0^2$ is the rotational diffusion coefficient measured previously, and $\xi_\delta$ and $\xi_\phi$ are uncorrelated Wiener processes with zero mean and variance $t$. This defines the generalized potential
   $ V(\delta,\phi)=-\frac{2}{r_0^2}\frac{k}{d_{12}(\delta,\phi)} \,$.
    \label{eq:Vdeltaphi}
These dynamics are simulated using an equivalent agent-based model solved with molecular dynamics [SI Sec.~S3].
Both the temporal dynamics [Fig.~\ref{fig3}(a)] and the probability density [Fig.~\ref{fig3}(c)] from simulations agree with our experimental findings with  $k=83\, \kB T{\cdot}\si{\micro m}$ \cite{10.1021/acsnano.4c18078}. Namely, simulations show synchronization for relative phases of $(2n+1)\pi$, where $n$ is an integer [Fig.~\ref{fig3}(c)], confirming the validity of the model. Simulations show enhanced synchronicity for counter-rotors compared to co-rotors: we understand this qualitatively by noting that the configuration that minimizes coupling energy lies more often within the synchronisation region ($\frac{\pi}{2}<\phi< \frac{3\pi}{2}$ or $\frac{\pi}{2}<\delta< \frac{3\pi}{2}$, respectively) for counter-rotors than for co-rotors, favoring phase-locking in the former case. We gain analytical insight by reducing  Eqs.~\eqref{eq:deltadotgen} and \eqref{eq:phidotgen} using separation of timescales. 
For co-rotors, $\delta$ varies slowly, as $\dot\delta\propto \omega_2-\omega_1$, while $\phi$ varies fast, as $\dot\phi\propto \omega_1+\omega_2 = \omega_\phi$. 
A period $T = 4\pi/|\omega_\phi|$ naturally arises, over which Eq.~\eqref{eq:deltadotgen} can be averaged, under the assumption that $\phi$ varies much faster than and independently from $\delta$. To leading order in $r_0/R \to 0$, this gives 
$
    \dot{\delta} \simeq -\frac{2D_r}{\kB T} \frac{\partial V_\delta}{\partial \delta}  + 2\sqrt{D_r}\,\xi_\delta
    \label{eq:deltadot_avg}  
$, 
where $V_\delta$ is the approximate effective potential
\begin{equation}
    V_\delta(\delta)  =  \frac{k_\mathrm{B}T}{2D_r}\left[-(\omega_2-\omega_1)\delta + \frac{\mu_0 k}{8 R^3} \cos\delta\right]\, 
    \label{eq:Vdelta}
\end{equation}
Full analytical calculations and numerical validation of the approximations for Eq.~\eqref{eq:Vdelta} are presented in SI Sec.~S2. This is a classical Adler model of synchronization~\cite{adler, Shlomovitz:2014gv, Aubret:2018cca, eLife.Brumley.2014}. The washboard potential  $V_\delta (\delta)$ 
presents local minima, which result from the periodicity of the cosine function and define synchronous states, separated by slips as observed in the experiment and simulations.
For small $\omega_2-\omega_1$, the position of the minima is given by $\delta_\mathrm{min} \approx (2k+1)\pi$, with $k$ integer [Fig.~\ref{fig3}(b-c)].
Eq.~\eqref{eq:Vdelta} only admits minima for $\delta$ if $|\omega_2-\omega_1| < \frac{\mu_0 k}{8 R^3}$, setting a necessary condition for the synchronization for co-rotors.
For counter-rotors, for which $\omega_1$ and $\omega_2$ have opposite signs, we obtain  an analogous equation for $\dot{\phi}$ 
 with  effective potential 
   $ V_\phi(\phi) \approx \frac{k_\mathrm{B}T}{2D_r}\left[-(\omega_1+\omega_2)\phi + 0.5\, \frac{\mu_0 k}{R^3} \cos\phi\right]$ [SI Sec.~S2]. 
The condition for the synchronization of two counter-rotors becomes $|\omega_1+\omega_2| < 0.5\, \frac{\mu_0 k}{R^3}$, less restrictive than  for co-rotors, 
in line with our intuition of why counter-rotors synchronise better than co-rotors in simulations, all other parameters being equal.

Taking co-rotors as example, we interpret the the duration $t_\delta$ of a synchronization plateau as a Kramers escape time from the energy barrier defined by the effective potential \eqref{eq:Vdelta}, reminiscent of stochastic switching between synchrony and asynchrony for swimming algae \cite{polin2009chlamydomonas,Wan2014}. When the angular frequencies are close, the energy barrier is approximately $\Delta V_\delta = V_\delta(\delta_\mathrm{max}) - V_\delta(\delta_\mathrm{min}) \simeq \frac{k_\mathrm{B}T}{2D_r}\left(\frac{\mu_0k}{4R^3} - \pi\,\Delta\omega\right)$, where $\delta_\mathrm{max}$ and $\delta_\mathrm{min}$ are a maximum and its consecutive minimum of $V_\delta(\delta)$,  and $\Delta \omega = |\omega_2-\omega_1|$ [SI Sec.~S2]. This results in a Kramers time
    $\tau_\delta = \frac{2\pi k_\mathrm{B}T}{(2D_r)\sqrt{V''_\delta(\delta_\mathrm{min})|V''_\delta(\delta_\mathrm{max})|}}\  \operatorname{e}^{\frac{\Delta V_\delta}{k_\mathrm{B}T}}\,\propto \operatorname{e}^{-\frac{\pi\Delta \omega}{2D_r}}$,
    \label{eq:kramers-time} 
intuitively related to the experimental synchronization ratio $\Gamma$. Guided by this theoretical insight, we represent the experimental synchronization $\Gamma$ vs $\exp{(-\lambda)}$, where $\lambda = \pi\frac{\Delta\omega}{2D_r}$ and observe a collapse of all our data: clocks of different radii $R$, co- or counter-rotating [Fig.~4b]. Similarly, the synchronization probability $p_\mathrm{sync}$  from our agent-based model [Fig. 4c-inset] collapses across all parameters varied numerically (coupling strength or radii  and velocity) when represented as a function of $\lambda$ [Fig.~4c] -- altogether  showing that population noise, {\it i.e.} frequency difference $\Delta \omega$, is counteracted by active noise $D_r$ to hinder synchronization.
{\bf Conclusion.}
We leveraged tools of active matter to devise a model platform of mechanical oscillators at the microscale, coupled phoretically, rather than hydrodynamically, whereby individual dynamics, natural frequency, noise and phoretic coupling are independently quantified. We observe synchronicity of pairs of oscillators separated by slips and controlled quantitatively by a parameter $\lambda\sim \Delta \omega/D_R $, expressed in terms of source of noise in the system.  It is noticeable that while we predict that counter-rotors synchronize more robustly than co-rotors, this is not observed in the experiment, pointing out to  
potential chemical shadowing and geometric effect that will require further investigation. This work lays the  groundwork  for the  study of synchronization of oscillators spatially and temporally coupled by non-reciprocal interactions. \\
{\bf Acknowledgements.} This research was funded in whole or in part by the Austrian Science Fund (FWF) [10.55776/P35206].
{\bf Conflicts of interest.} No conflict to declare.




\providecommand*{\mcitethebibliography}{\thebibliography}
\csname @ifundefined\endcsname{endmcitethebibliography}
{\let\endmcitethebibliography\endthebibliography}{}

\bibliographystyle{rsc} 

\nocite{Paquay2016,Thompson2022,Stukowski2009}


\end{document}